\documentclass[fleqn,a4paper,11pt,dvips]{article}
\usepackage[margin=1in]{geometry}
\usepackage{colortbl, subfigure}
\usepackage{amssymb, amsmath, enumerate}
\usepackage{amsthm}
\usepackage{mathrsfs}
\usepackage[linesnumbered, ruled]{algorithm2e}
\usepackage{cite}
\usepackage{color}

\newtheorem{theorem}{Theorem}[section]
\newtheorem{lemma}[theorem]{Lemma}

\newtheorem{corollary}[theorem]{Corollary}

\newtheorem{definition}[theorem]{Definition}

\newcommand{\doublespace} {\addtolength{\baselineskip}{0.05\baselineskip}}

\newif\ifpdf
  \ifx\pdfoutput\undefined
  \pdffalse
\else
  \pdfoutput=1
  \pdftrue
\fi

\ifpdf
  \usepackage[pdftex]{graphicx}
\else
  \usepackage{graphicx}
\fi

\def\qed{\hfill $\Box$\vspace{0.3cm}}
\def\pf{\noindent{\bf Proof. }}
\title{\LARGE\bf On the Orbits of Crossed Cubes}
\author{
\small{Tzong--Huei Shiau$^{1}$, Yue--Li Wang$^{2,}$\thanks{All correspondence should be
addressed to Professor Yue-Li Wang, Department of Information
Management, National Taiwan University of Science and Technology,
43, Section 4, Kee-Lung Road, Taipei, Taiwan 10607 (e-mail:
ylwang@cs.ntust.edu.tw).},\ \ and \ \ Kung--Jui Pai$^{3}$}
\\
\\
{\small $^1$ Computer \& Communications Associates Incorporation, Hinchu, Taiwan}\\
{\small $^2$ Department of Information Management,}\\
{\small National Taiwan University of Science and Technology, Taipei, Taiwan}\\
{\small $^{3}$ Department of Industrial Engineering and Management,}\\
{\small  Ming Chi University of Technology, New Taipei City, Taiwan}\\ \\
}
\date{}

\begin{document}

\doublespace \baselineskip=24pt \maketitle \thispagestyle{empty}

\begin{abstract} \normalsize
An orbit of $G$ is a subset $S$ of $V(G)$ such that $\phi(u)=v$ for any two vertices $u,v\in S$,
where $\phi$ is  an isomorphism of $G$.
The orbit number of a graph $G$, denoted by $\text{Orb}(G)$, is the number of
orbits of $G$. In [A Note on Path Embedding in Crossed Cubes with Faulty Vertices, 
Information Processing Letters 121 (2017) pp. 34--38], Chen et al. conjectured that $\text{Orb}(\text{CQ}_n)=2^{\lceil\frac{n}{2}\rceil-2}$ for $n\geqslant 3$, where $\text{CQ}_n$ denotes an $n$-dimensional crossed cube.
In this paper, we settle the conjecture.
\vskip 0.4in

\noindent{\bf Keywords:} Crossed cubes; Automorphism; Vertex-transitive;  Orbits.
\end{abstract}

\setcounter{page}{0}
\newpage
\section{\hspace{-0.5cm}. Introduction}

Let $G=(V,E)$ be a graph with vertex set $V(G)$ and edge set $E(G)$ which are simply
denoted by $V$ and $E$, respectively, when the context is clear.
An {\em automorphism} of a graph $G=(V,E)$ is a mapping $\phi: V(G)\rightarrow V(G)$
such that there is an edge $uv\in E(G)$ if and only if  $\phi(u)\phi(v)$  is also an edge in $E(G)$.
A graph is {\em vertex-transitive} if, for any two vertices $u$ and $v$ of $G$,
there is an automorphism $\phi$ such that $\phi(u)=v$. Clearly, every vertex-transitive graph is regular.
However, not all regular graphs are vertex-transitive, e.g., crossed cubes \cite{Efe91,Efe92} and the Frucht graph \cite{Fruc39}.

\begin{definition}\label{def:orbits}
{\em
An orbit of $G$ is a subset $S$ of $V(G)$ such that $\phi(u)=v$ for any two vertices $u,v\in S$,
where $\phi$ is  an isomorphism of $G$.
The {\em orbit number} of a graph $G$, denoted by $\text{Orb}(G)$, is the number of
orbits in $G$.}
\end{definition}

By Definition~\ref{def:orbits}, all vertex-transitive graphs $G$ are with $\text{Orb}(G)=1$, e.g. hypercubes.
In \cite{Efe91,Efe92}, Efe introduced the crossed cubes which will be defined in Section~\ref{Preliminaries}. 
Crossed cubes have several properties, e.g., smaller diameter and better embedding properties, which makes it compare favorably
to the ordinary hypercubes \cite{Efe92,Saad88,Seit85}. The crossed cubes have been extensively studied \cite{Chan00,Chan14,Chen15,Chen12,Chen17,Chen13a,Chen13b,Dong10,Dong12,Efe91,Efe92,Efe94,Fan02,Fan05,Fan07,Fu09,Kula95,Kula95b,Kula97,Ma08,Park07,Tsai15,Wang12,Xu06,Yang03,Yang10,Zhan13,Zhen96,Zhou10}.
In \cite{Kula95}, Kulasinghe and Bettayeb showed that $\text{Orb}(\text{CQ}_n)>1$ when $n\geqslant 5$, where
$\text{CQ}_n$ is the $n$-dimensional crossed cube. In \cite{Chen17}, Chen et al. showed that $\text{Orb}(\text{CQ}_5)=2$ and conjectured that  $\text{Orb}(\text{CQ}_n)=2^{\lceil\frac{n}{2}\rceil-2}$ for $n\geqslant 3$.
In this paper, we settle the conjecture.

The rest of this paper is organized as follows. In
Section~\ref{Preliminaries}, we introduce the definition of crossed cubes. In
Section~\ref{UpperBound}, we show that $2^{\lceil\frac{n}{2}\rceil-2}$ is an upper
bound of $\text{Orb}(\text{CQ}_n)$ for $n\geqslant 3$.
In Section~\ref{LowerBound}, we show that $2^{\lceil\frac{n}{2}\rceil-2}$ is also a lower
bound of $\text{Orb}(\text{CQ}_n)$ for $n\geqslant 3$.
Finally, Section~\ref{conclusion} contains our
concluding remarks.

\section{\hspace{-0.5cm}. Preliminaries}
\label{Preliminaries}

\begin{definition}
Two $2$-bit binary strings $x_2x_1$ and $y_2y_1$ are pair related, denoted by $x_2x_1 \sim y_2y_1,$ if and only if $(x_2x_1, y_2y_1) \in \{(00, 00), (10, 10), (01, 11), (11, 01)\}$.
\end{definition}

The crossed cubes were introduced by Efe in \cite{Efe91,Efe92}. The $n$-dimensional crossed cubes $\text{CQ}_n$
contains $2^n$ vertices in which the degree of every vertex is $n$.
Every vertex of $\text{CQ}_n$ is identified by a unique binary string,
which is also called {\em address}, of length $n$.
Let $u=u_{n-1}\ldots u_0$  be a vertex in $V(\text{CQ}_n)$.
A vertex $u$ is an {\em even vertex} (respectively, {\em odd vertex}) if the value of $u_{n-1}\ldots u_0$ is even (respectively, odd).
The {\em negate} of $u_i$ for $0\leqslant i\leqslant n-1$ is denoted by $\overline{u}_i$.
For an index $x$ with $0\leqslant x\leqslant n-1$, we use $P^x_u$ to denote the prefix $u_{n-1}\ldots u_{x+1}$ while
$S^x_u$ denotes the suffix $u_{x-2}\ldots u_0$ (respectively, $u_{x-1}\ldots u_0$) when $x$ is odd (respectively, even).
If $u_i=v_i$ for all $x+1\leqslant i \leqslant n-1$, then we use $P^x_u=P^x_v$ to denote it.
Furthermore, let $S^x_u\sim S^x_v$ stand for $u_{2i+1}u_{2i}\sim v_{2i+1}v_{2i}$ for all $0\leqslant i \leqslant \lfloor\frac{x}{2}\rfloor-1$.

The edges of an $n$-dimensional crossed cube can be defined as follows.

\begin{definition}\label{EdgesinCQn}
Let $u=u_{n-1}\ldots u_0$ and $v=v_{n-1}\ldots v_0$ be two vertices in $\text{CQ}_n$.
There is an edge $uv$ in $E(\text{CQ}_n)$ if and only if there exists an index $x$ with $0\leqslant x\leqslant n-1$ such that
the following conditions are satisfied
\begin{enumerate}
\item $v_x=\overline{u}_x$
\item $v_{x-1}=u_{x-1}$ if $x$ is odd,
\item $P^x_v=P^x_u$, and
\item $S^x_v\sim S^x_u$.
\end{enumerate}
\end{definition}

We say that vertex $v$ is the {\em $k$th-neighbor} of vertex $u$ if $u$ and $v$ are adjacent along dimension $k$.
That is, $v_k=\overline{u}_k$, $v_{k-1}=u_{k-1}$ if $k$ is odd, $P^k_v=P^k_u$, and $S^k_v\sim S^k_u$.

\noindent {\bf Example 1.} Figure 1 depicts $\text{CQ}_3$ and $\text{CQ}_4$. For example, let $u=0011$ and $v=0101$.
We can find that $v_2=\overline{u}_2$, $P^2_v=P^2_u=0$, and $S^2_v=01\sim 11=S^2_u$. Thus there is an edge $uv$ in $E(\text{CQ}_4)$.\\

\begin{figure}[htb]
\begin{center}
\subfigure[CQ$_3$]{
\includegraphics[scale=0.28]{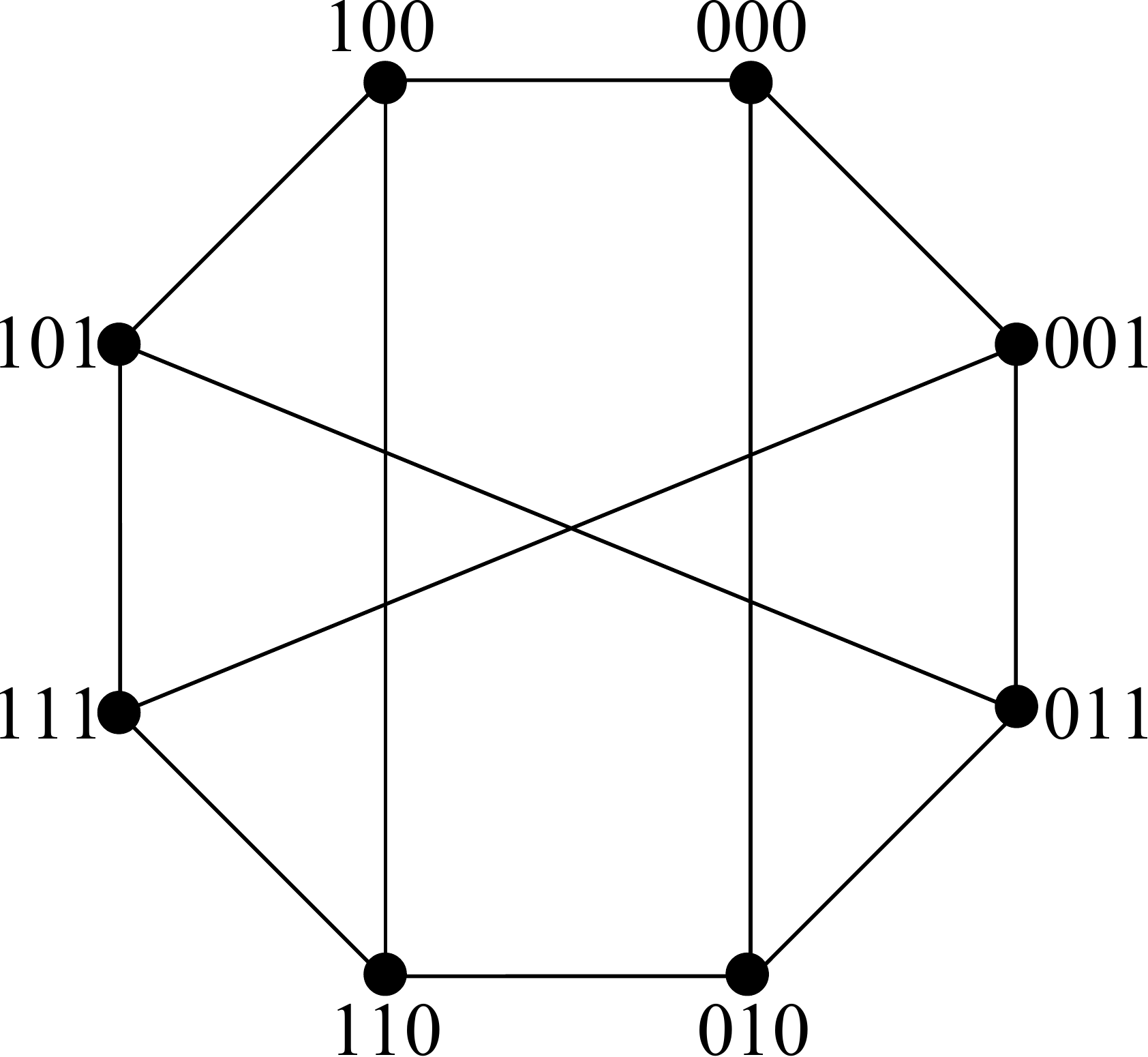}
} \quad \subfigure[CQ$_4$]{
\includegraphics[scale=0.15]{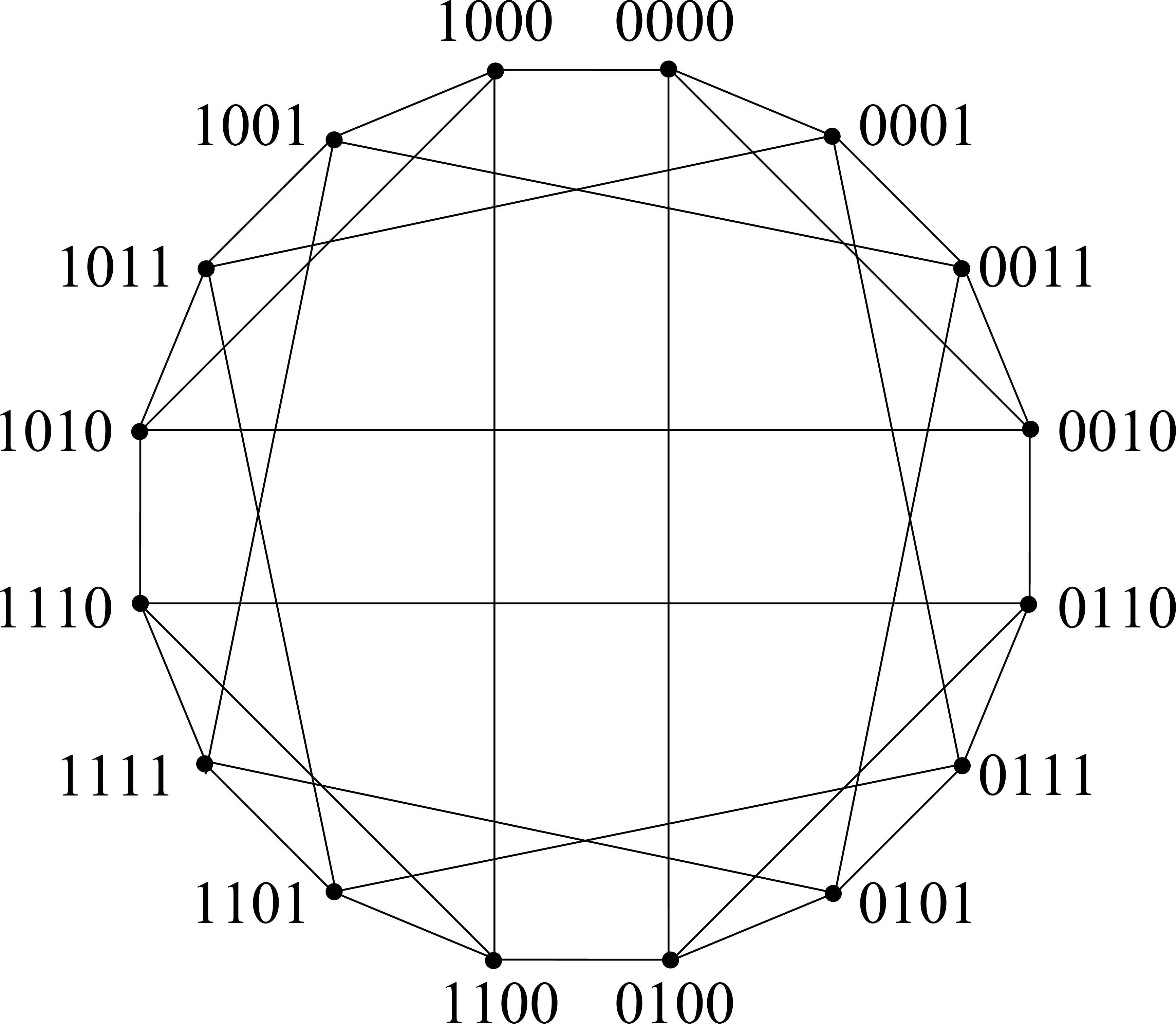}
}  \caption{\label{fig1}Crossed cubes $\text{CQ}_3$ and $\text{CQ}_4$.}
\end{center}
\end{figure}

As far as we know, the following theorem is the only result on the orbit number of crossed cubes.

\begin{theorem}[\hspace{-0.015cm}\cite{Kula95}]\label{thm:theore1}
$Orb(\text{CQ}_n)>1$ when $n > 4$.
\end{theorem}

\section{\hspace{-0.5cm}. The upper bound of $Orb(\text{CQ}_n)$}
\label{UpperBound}

In this section, we show that $\text{Orb}(\text{CQ}_n)\leqslant 2^{\lceil\frac{n}{2}\rceil-2}$ for $n\geqslant 3$.
In the rest of this paper, we use $f_i(u)$ to denote $u_{n-1}\ldots u_{i+1}\overline{u}_i u_{i-1}\ldots u_0$
for some $0\leqslant i\leqslant n-1$.

\begin{lemma}\label{lm:lemma1}
For $n\geqslant 2$ and any odd $k$ with $1\leqslant k <n$, the function $\phi(u)=f_k(u)$ for all vertices $u\in V(\text{CQ}_n)$ is an automorphism of $\text{CQ}_n$.
\end{lemma}
\pf
By the definition of automorphisms, we have to show that, for each edge $uv\in E(\text{CQ}_n)$, there is an edge $\phi(u)\phi(v)\in E(\text{CQ}_n)$.
By Definition~\ref{EdgesinCQn}, we have that $P^x_v=P^x_u$, $v_x=\overline{u}_x$, and $S^x_u\sim S^x_v$ for some $0\leqslant x\leqslant n-1$.
Moreover, $v_{x-1}=u_{x-1}$ when $x$ is odd.
For simplicity, we only consider the case where $x$ is odd. The other case can be handled similarly. We distinguish the following three cases.

\noindent {\bf Case 1.} $k<x$.

In this case, we have that $\phi(u)=P_u^x u_xu_{x-1}\ldots\overline{u}_k u_{k-1}S^k_u$ and $\phi(v)=P_v^x v_xv_{x-1}\ldots\overline{v}_k v_{k-1}S^k_v$.
Since only the $k$th bit is changed in $S^x_u$ and $S^x_v$,
all we have to prove is that $\overline{u}_k u_{k-1}\sim \overline{v}_k v_{k-1}$. Table~\ref{tbl:Table1} lists
all possible cases of $\overline{u}_k u_{k-1}$ and $\overline{v}_k v_{k-1}$. It is easy to check that $\overline{u}_k u_{k-1}\sim \overline{v}_k v_{k-1}$
for any case. Thus there is an edge $\phi(u)\phi(v)$ in $E(\text{CQ}_n)$ and this case holds.

\begin{table}[htb]
\begin{center}
\caption{\label{tbl:Table1}All possible cases of $\overline{u}_k u_{k-1}$ and $\overline{v}_k v_{k-1}$}
\begin{tabular}{|c|c|c|c|}
\hline {$u_{k}u_{k-1}$} & {$\overline{u}_{k}u_{k-1}$}& {$v_{k}v_{k-1}$} & {$\overline{v}_{k}v_{k-1}$} \\
\hline
\hline 00 & {10} & {00} & {10} \\
\hline 01 & {11} & {11} & {01} \\
\hline 10 & {00} & {10} & {00} \\
\hline 11 & {01} & {01} & {11} \\
\hline
\end{tabular}
\end{center}
\end{table}

\noindent {\bf Case 2.} $k=x$.

In this case, we have that $\phi(u)=P_u^x\overline{u}_x u_{x-1}S^x_u$ and $\phi(v)=P_v^x\overline{v}_x v_{x-1}S^x_v$.
It is clear that $P_u^x=P_v^x$, $u_{x-1}=v_{x-1}$, and $S^x_u\sim S^x_v$.
Note that $\overline{v}_x=\overline{\overline{u}}_x=u_x$. By Definition~\ref{EdgesinCQn}, there is an edge $\phi(u)\phi(v)$ in $E(\text{CQ}_n)$.
Thus this case also holds.

\noindent {\bf Case 3.} $k>x$.

In this case, only a bit in the same position of $P_u^x$ and $P_v^x$ is negated. Thus
$P_{\phi(u)}^x=P_{\phi(v)}^x$ and all the other relations between $u_x$ and $v_x$, $u_{x-1}$ and $v_{x-1}$, and $S^x_u$ and $S^x_v$
remain unchanged. By Definition~\ref{EdgesinCQn}, there is an edge $\phi(u)\phi(v)$ in $E(\text{CQ}_n)$. This completes the proof.
\qed

\begin{lemma}\label{lm:lemma2}
For $n\geqslant 2$, the function $\phi(u)=f_{n-1}(u)$ for all vertices $u\in V(\text{CQ}_n)$  is an automorphism of $\text{CQ}_n$.
\end{lemma}
\pf
If $n$ is even, then $n-1$ is an odd number.
By Lemma~\ref{lm:lemma1}, the function $\phi$ is an automorphism of $\text{CQ}_n$.

Now we consider the case where
$n$ is odd. Assume that there is an edge $uv\in E(\text{CQ}_n)$ with $P_u^k=P_v^k$, $S_v^k\sim S_u^k$, and $v_k=\overline{u}_k$ (and $v_{k-1}=u_{k-1}$
when $k$ is odd). If $k\neq n-1$, then, by using a similar argument as in Lemma~\ref{lm:lemma1},
we can prove that there is also an edge $\phi(u)\phi(v)\in E(\text{CQ}_n)$.
It remains to consider the case where $k=n-1$ and $n$ is odd.
Accordingly,
we have $\phi(u)=\overline{u}_{n-1}S_u^{n-1}$ and $\phi(v)=\overline{v}_{n-1}S_v^{n-1}$.
It is easy to verify that $\phi(u)\phi(v)$ is also an edge in $E(\text{CQ}_n)$.
This completes the proof.
\qed

\begin{lemma}\label{lm:lemma3}
For $n\geqslant 2$, the function $\phi(u)=f_{n-2}(u)$ for all vertices $u\in V(\text{CQ}_n)$ is an automorphism of $\text{CQ}_n$.
\end{lemma}
\pf
If $n$ is odd, then $n-2$ is an odd number.
By Lemma~\ref{lm:lemma1}, the function $\phi$ is an automorphism of $\text{CQ}_n$.

Now we consider the case where
$n$ is even. Assume that there is an edge $uv\in E(\text{CQ}_n)$ with $P_u^k=P_v^k$, $S_v^k\sim S_u^k$, $v_k=\overline{u}_k$ and (and $v_{k-1}=u_{k-1}$
when $k$ is odd). If $k\leqslant n-2$, then, by using a similar argument as in Lemma~\ref{lm:lemma2},
we can prove that there is also an edge $\phi(u)\phi(v)\in E(\text{CQ}_n)$.
It remains to consider the case where $k=n-1$.
Accordingly, we have Table~\ref{tbl:Table2}. It is easy to check from Table~\ref{tbl:Table2} that there is an edge $\phi(u)\phi(v)\in E(\text{CQ}_n)$.
This completes the proof. \qed

\begin{table}[htb]
\begin{center}
\caption{\label{tbl:Table2}All possible cases of $u_{n-1} \overline{u}_{n-2}$ and $v_{n-1} \overline{v}_{n-2}$}
\begin{tabular}{|c|c|c|c|}
\hline {$u_{n-1}u_{n-2}$} & {$v_{n-1}v_{n-2}$}& {$u_{n-1} \overline{u}_{n-2}$} & {$v_{n-1} \overline{v}_{n-2}$} \\
\hline
\hline 00 & {10} & {01} & {11} \\
\hline 01 & {11} & {00} & {10} \\
\hline 10 & {00} & {11} & {01} \\
\hline 11 & {01} & {10} & {00} \\
\hline
\end{tabular}
\end{center}
\end{table}

\begin{lemma}\label{lm:lemma4}
For odd $n\geqslant 3$ and $u\in V(\text{CQ}_n)$, the following function $\phi$ is an automorphism of $\text{CQ}_n$:
\begin{equation*}
\phi(u) =\begin{cases}
f_{n-3}(u) & \quad \text{if $u_{n-1}=0$} \\
f_{n-3}(f_{n-2}(u)) & \quad \text{otherwise.}
\end{cases}
\end{equation*}
\end{lemma}
\pf It is easy to show that the function $\phi$ is a bijective function.
It remains to prove that $\phi$ is an automorphism of $\text{CQ}_n$.
Let $uv$ be an edge in $E(\text{CQ}_n)$ with $P_u^k=P_v^k, u_k=\overline{v}_k$, and $S_u^k=S_v^k$
for some $0\leqslant k\leqslant n-1$. We claim that $\phi(u)\phi(v)$ is also
an edge in $E(\text{CQ}_n)$. If $k<n-3$, then it is clear that $\phi(u)\phi(v)$ is also
an edge in $E(\text{CQ}_n)$. Thus we consider the cases where $k=n-1,n-2$, and $n-3$.
For the case where $k=n-1$, if $u_{n-1}=0$ (respectively, $u_{n-1}=1$), then $v_{n-1}=1$ (respectively, $v_{n-1}=0$)
and $u_{n-2}u_{n-3}\sim v_{n-2}v_{n-3}$.
After the mapping of $\phi$, we can find that $\phi(u)_{n-1}=0$, $\phi(v)_{n-1}=1$ (respectively, $\phi(u)_{n-1}=1$ and $\phi(v)_{n-1}=0$), and
$\phi(u)_{n-2}\phi(u)_{n-3}\sim \phi(v)_{n-2}\phi(v)_{n-3}$ (see the first, second, fifth, and
sixth columns in Table~\ref{tbl:Table3}).
By using a similar argument, we can show that the claim holds for the other cases. This completes the proof.
\qed

\begin{table}[htb]
\begin{center}\small
\caption{\label{tbl:Table3}The leftmost three bits of $u$, $v$, $\phi(u)$, and $\phi(v)$}
\begin{tabular}{|c|c|c|c|c|c|c|c|}\hline
\multicolumn{1}{|c|}{$u$} & \multicolumn{3}{|c|}{$v$ with $k=$} &
\multicolumn{1}{|c|}{$\phi(u)$} & \multicolumn{3}{|c|}{$\phi(v)$ with $k=$} \\
[2pt]\hline\hline  & $n-1$ & $n-2$ & $n-3$ & & $n-1$ & $n-2$ & $n-3$\\
\hline
000& 100 & 010 & 001 & 001 & 111 & 011 & 000 \\
001& 111 & 011 & 000 & 000 & 100 & 010 & 001 \\
010& 110 & 000 & 011 & 011 & 101 & 001 & 010 \\
011& 101 & 001 & 010 & 010 & 110 & 000 & 011 \\
100& 000 & 110 & 101 & 111 & 001 & 101 & 110 \\
101& 011 & 111 & 100 & 110 & 010 & 100 & 111 \\
110& 010 & 100 & 111 & 101 & 011 & 111 & 100 \\
111& 001 & 101 & 110 & 100 & 000 & 110 & 101 \\
\hline
\end{tabular}
\end{center}
\end{table}

\newpage
\begin{lemma}\label{lm:lemma5}
For even $n\geqslant 4$ and $u\in V(\text{CQ}_n)$, the following function $\phi$ is an automorphism of $\text{CQ}_n$:
\begin{equation*}
\phi(u) =\begin{cases}
f_{n-4}(f_{n-3}(u)) & \quad \text{if $u_{n-1}u_{n-2}u_{n-3}u_{n-4}\in\{0100,1000, 0111,1011\}$}\\
f_{n-4}(u)& \quad \text{otherwise.}
\end{cases}
\end{equation*}
\end{lemma}
\pf By using a similar argument as in Lemma~\ref{lm:lemma4}, this lemma holds.
\qed

\begin{lemma}\label{lm:lemma6}
Let $\phi$ be an automorphism defined in Lemmas~\ref{lm:lemma1}-\ref{lm:lemma5}.
If $\phi(u)=v$, then $\phi(v)=u$.
\end{lemma}
\pf It is clear that the lemma holds for the $\phi$ defined in Lemmas~\ref{lm:lemma1}-\ref{lm:lemma3}.
For the case where $\phi$ is defined in Lemma~\ref{lm:lemma4} (respectively, Lemma~\ref{lm:lemma5}),
it is easy to check that $u=f_{n-2}(f_{n-2}(u))$ (respectively, $u=f_{n-3}(f_{n-3}(u))$) for $u$ with $u_{n-1}\neq 0$
(respectively, $u_{n-1}u_{n-2}u_{n-3}u_{n-4}\in\{0100,1000, 0111,1011\}$). This further implies that
the lemma holds for the $\phi$ defined in Lemmas~\ref{lm:lemma4} and \ref{lm:lemma5}. This completes the proof.
\qed

\begin{corollary}\label{coro:corollary1}
Let $\phi$ be an automorphism defined in Lemmas~\ref{lm:lemma1}-~\ref{lm:lemma5}.
Every orbit contains exactly two vertices under the automorphism $\phi$.
\end{corollary}

\begin{lemma}\label{lm:lemma7}
For $n\geqslant 2$ and $k$ even, the function $\phi(u)=f_k(u)$ for all vertices $u\in V(\text{CQ}_n)$  is an automorphism of $\text{CQ}_n$ only when $k$ is in $\{n-2, n-1\}$.
\end{lemma}
\pf Note that if $k$ is even, then $k=n-2$ when $n$ is even and $k=n-1$ when $n$ is odd.
If $k$ is in $\{n-2, n-1\}$, then, by using
a similar argument as in Lemma~\ref{lm:lemma1}, it is easy to show that
negating the $k$th  bit is an automorphism of $\text{CQ}_n$. It remains to show that
it is impossible to find an automorphism of $\text{CQ}_n$ by negating the $k$th bit
of the addresses of all vertices in $\text{CQ}_n$ when $k$ is even and $k\notin \{n-2, n-1\}$.
It is clear that there exists an edge $uv$ in $E(\text{CQ}_n)$ with $k+1<x$ such that
$P^x_v=P^x_u$, $v_x=\overline{u}_x$, and $S^x_v\sim S^x_u$ when $k\notin \{n-2, n-1\}$.
By examining all possible cases of $u_{k+1}\overline{u}_k$ and $v_{k+1}\overline{v}_k$ (see Table~\ref{tbl:Table4}),
the relation of $u_{k+1}\overline{u}_k\sim v_{k+1}\overline{v}_k$ does not exist.
Thus there is no edge between $\phi(u)$ and $\phi(v)$. This completes the proof.
\qed

\begin{table}[htb]
\begin{center}
\caption{\label{tbl:Table4}All possible cases of $u_{k+1} \overline{u}_k$ and $v_{k+1}\overline{v}_k$}
\begin{tabular}{|c|c|c|c|}
\hline {$u_{k+1}u_k$} & {$u_{k+1} \overline{u}_k$}& {$v_{k+1}v_k$} & {$v_{k+1}\overline{v}_k$} \\
\hline
\hline 00 & {01} & {00} & {01} \\
\hline 01 & {00} & {11} & {10} \\
\hline 10 & {11} & {10} & {11} \\
\hline 11 & {10} & {01} & {00} \\
\hline
\end{tabular}
\end{center}
\end{table}

\begin{lemma}\label{lm:lemma8}
For $n\geqslant 3$, $\text{Orb}(\text{CQ}_n)\leqslant 2^{\lceil\frac{n}{2}\rceil-2}$.
\end{lemma}
\pf By Corollary~\ref{coro:corollary1}, every orbit contains exactly two vertices under an automorphism $\phi$
defined in Lemmas~\ref{lm:lemma1}-\ref{lm:lemma5}.
Note that, after applying two different automorphisms, we have that each orbit contains four distinct vertices.
Since there are $\lfloor\frac{n}{2}\rfloor+2$ different automorphisms defined in Lemmas~\ref{lm:lemma1}-\ref{lm:lemma5},
it follows that each orbit contains $2^{\lfloor\frac{n}{2}\rfloor+2}$ distinct vertices.
This further implies that $\text{Orb}(\text{CQ}_n)\leqslant \frac{2^n}{2^{\lfloor\frac{n}{2}\rfloor+2}}=2^{\lceil\frac{n}{2}\rceil-2}$.
This completes the proof.
\qed

\begin{corollary}\label{coro:corollary2}
$\text{Orb}(\text{CQ}_3)=\text{Orb}(\text{CQ}_4)=1$ and $\text{Orb}(\text{CQ}_5)=\text{Orb}(\text{CQ}_6)=2$.
\end{corollary}

\section{\hspace{-0.5cm}. The lower bound of $Orb(\text{CQ}_n)$}
\label{LowerBound}

Denote by $N(v) = \{u\in V: uv\in E\}$ the {\em open neighborhood}
of $v$. A path (respectively, clique) in $G$ of $\ell$ vertices is denoted by $P_\ell$ (respectively, $K_\ell$).

\begin{definition}
The {\em $P_4$-graph} of a set $S\subseteq V$
is a graph $H$ with $V(H)=S$ and there is an edge $xy$ for $x,y\in V(H)$ if and only
if there is a $P_4$ from $x$ to $y$ in $G$.
\end{definition}

\begin{figure}[htb]
\begin{center}
\subfigure[the $P_4$-graph of $N(1)$]{
\includegraphics[scale=0.3]{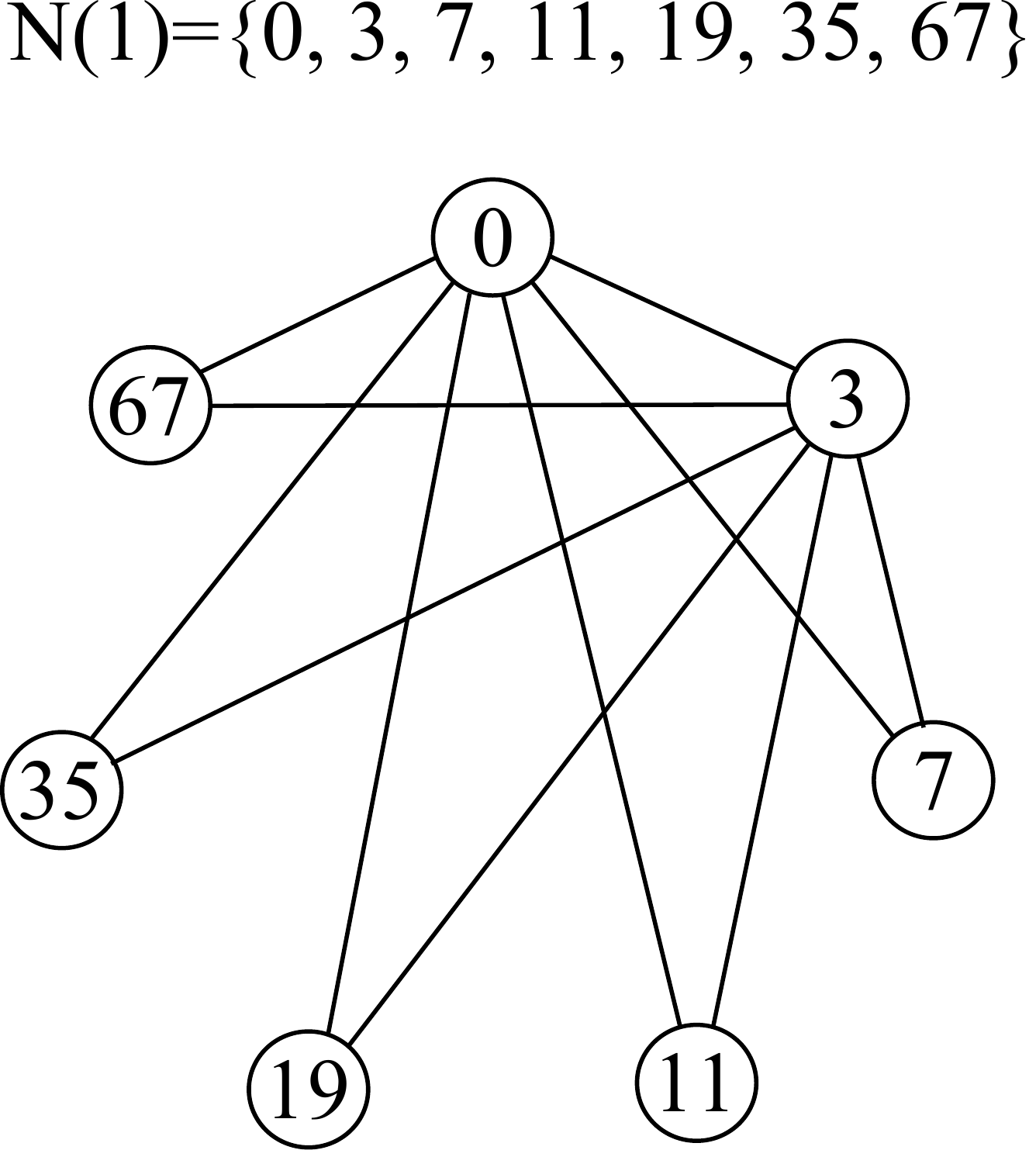}
} \quad \subfigure[the $P_4$-graph of $N(4)$]{
\includegraphics[scale=0.3]{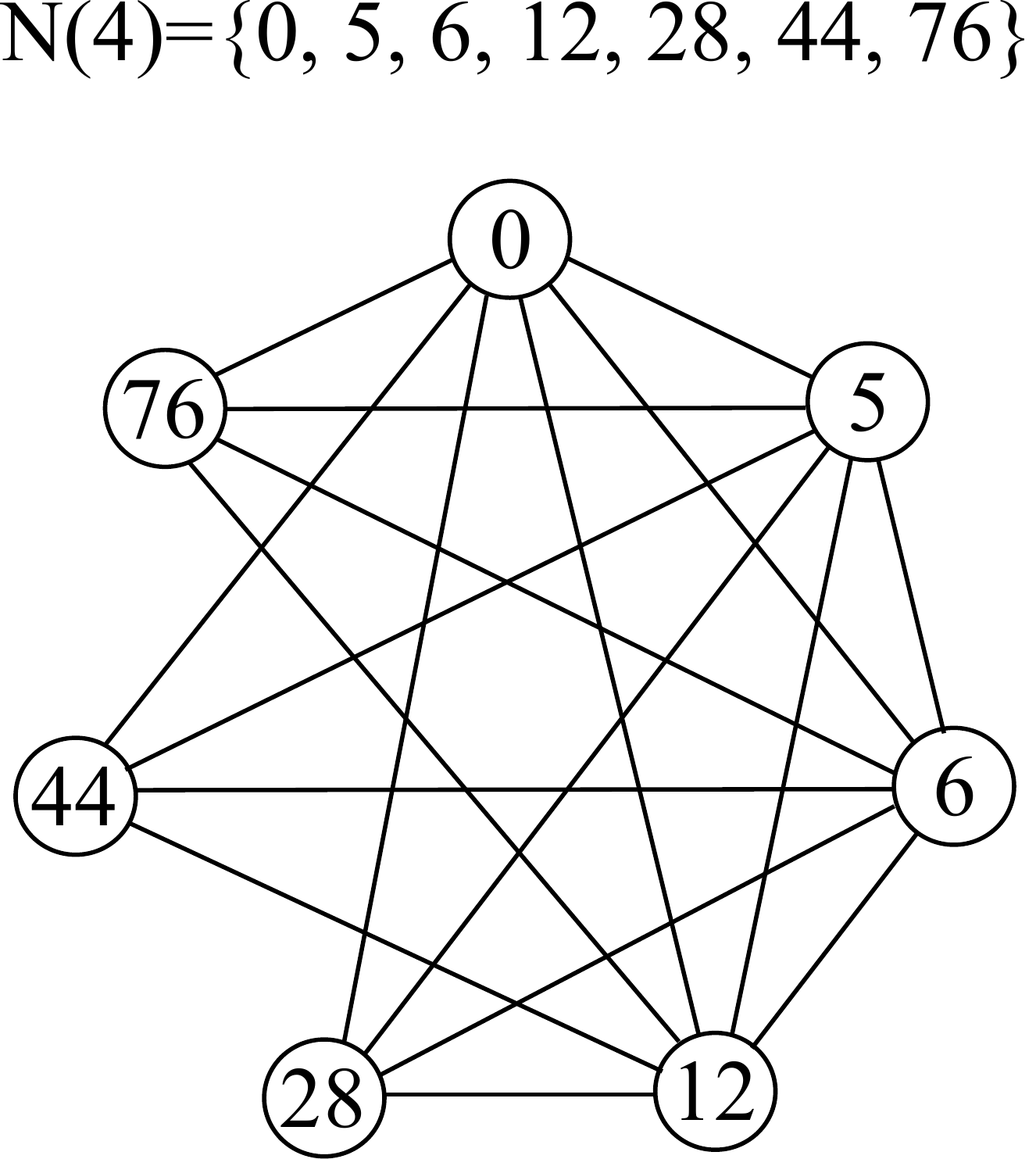}
}  \caption{\label{fig2}The $P_4$-graphs of $N(1)$ and $N(4)$ in $\text{CQ}_7$.}
\end{center}
\end{figure}

\noindent {\bf Example 2.}
Figure~\ref{fig2} depicts the $P_4$-graphs of $N(1)$ and $N(4)$  in $\text{CQ}_7$.
We only explain the construction of Figure~\ref{fig2}(a).
First we show that there is a $P_4$ in $\text{CQ}_7$ from vertex $0$ to each
vertex in $N(1)\setminus \{0\}$. Since all vertices in $N(1)\setminus \{0,3\}$
have exactly three nonzero bits in their addresses, there is a $P_4$ in $\text{CQ}_7$ from vertex 0 to every
vertex in $N(1)\setminus \{0,3\}$. It is easy to check that the path $0, 1, 7, 3$ is also a $P_4$ in $\text{CQ}_7$ from vertex $0$ to vertex $3$.
Now we show that there is a $P_4$ in $\text{CQ}_7$ from vertex $3$ to each
vertex in $N(1)\setminus \{0,3\}$. It is easy to find a $P_3$ from vertex $2$ passing through
vertices $6$, $10$, $18$, $34$, and $66$ to vertices $7$, $11$, $19$, $35$, and $67$, respectively.
Since vertex $3$  is adjacent to vertex $2$. Thus there exists a $P_4$ from vertex $3$ to each
vertex in $N(1)\setminus \{0,3\}$. The reason why there is no $P_4$ between any two vertices
in $N(1)\setminus \{0,3\}$ will be explained in Lemma~\ref{lm:lemma12}.\\

\begin{lemma}\label{lm:lemma9}
For $n\geqslant 5$, if $u$ is an even vertex in $\text{CQ}_n$, then the $P_4$-graph of $N(u)$ contains a
$K_4$.
\end{lemma}
\pf Let $u=u_{n-1}\ldots u_0$  be an even vertex in $V(\text{CQ}_n)$.
It is obviously that $u_0=0$. We can find that $w=u_{n-1}\ldots u_1 1$,
$x=u_{n-1}\ldots \overline{u}_10$,
$y=u_{n-1}\ldots \overline{u}_2u_10$, and
$z=u_{n-1}\ldots \overline{u}_3u_2u_10$ are four vertices in $N(u)$.
It suffices to show that the induced subgraph of vertices $w, x, y$, and $z$ in
the $P_4$-graph of $N(u)$ is a $K_4$.
It is easy to check that the following five paths are $P_4$ in $\text{CQ}_n$:
$w=u_{n-1}\ldots u_2u_1 1\rightarrow u_{n-1}\ldots \overline{u}_2\overline{u}_1 1\rightarrow u_{n-1}\ldots \overline{u}_2\overline{u}_1 0\rightarrow u_{n-1}\ldots u_2\overline{u}_1 0=x$\\
$w=u_{n-1}\ldots u_2u_1 1\rightarrow u_{n-1}\ldots \overline{u}_2\overline{u}_1 1\rightarrow u_{n-1}\ldots \overline{u}_2\overline{u}_1 0\rightarrow u_{n-1}\ldots \overline{u}_2u_1 0=y$\\
$w=u_{n-1}\ldots u_2u_1 1\rightarrow u_{n-1}\ldots \overline{u}_3u_2\overline{u}_1 1\rightarrow u_{n-1}\ldots \overline{u}_3u_2\overline{u}_1 0\rightarrow u_{n-1}\ldots \overline{u}_3u_2u_1 0=z$\\
$x=u_{n-1}\ldots \overline{u}_10\rightarrow u_{n-1}\ldots u_2\overline{u}_11\rightarrow u_{n-1}\ldots \overline{u}_2u_11\rightarrow u_{n-1}\ldots \overline{u}_2u_10=y$, and\\
$x=u_{n-1}\ldots \overline{u}_10\rightarrow u_{n-1}\ldots u_2\overline{u}_11\rightarrow u_{n-1}\ldots \overline{u}_3u_2u_11\rightarrow u_{n-1}\ldots \overline{u}_3u_2u_10=z$.\\
It remains to show that there is a $P_4$ from $y$ to $z$ in $\text{CQ}_n$. If $u_2=0$, then the path\\
$y=u_{n-1}\ldots \overline{u}_30u_10\rightarrow u_{n-1}\ldots \overline{u}_31u_10\rightarrow u_{n-1}\ldots \overline{u}_4u_31u_10\rightarrow u_{n-1}\ldots u_4u_3\overline{u}_2u_10=z$\\
is a $P_4$ from $y$ to $z$; otherwise, the path\\
$y=u_{n-1}\ldots \overline{u}_31u_10\rightarrow u_{n-1}\ldots \overline{u}_4u_31u_10\rightarrow u_{n-1}\ldots \overline{u}_4u_30u_10\rightarrow u_{n-1}\ldots u_4u_3\overline{u}_2u_10=z$\\
is a $P_4$ from $y$ to $z$. This completes the proof.
\qed

\begin{lemma}[\hspace{-0.015cm}\cite{Kula95}]\label{lm:lemma10}
Let $v$ and $w$ be two vertices of $\text{CQ}_n$ such
that $v_{2k+1}v_{2k} = w_{2k+1}w_{2k} = 01$ for some $k$. Then $v$
and $w$ cannot be adjacent along a dimension greater than or equal to $2k$.
\end{lemma}

\begin{lemma}[\hspace{-0.015cm}\cite{Kula95}]\label{lm:lemma11}
Let $v$ and $w$ be two vertices in $\text{CQ}_n$ such
that $v_{2k+1}v_{2k} = 01$ and $w_{2k+1}w_{2k} = 10$ for some $k$.
Then $v$ and $w$ cannot be adjacent.
\end{lemma}

\begin{lemma}\label{lm:lemma12}
For $n\geqslant 5$, if $u$ is an odd vertex in $\text{CQ}_n$, then the $P_4$-graph of $N(u)$ contains no
$K_4$.
\end{lemma}
\pf If $u$ is an odd vertex, then there is exactly one even vertex in $N(u)$.
It suffices to show that, in the $P_4$-graph of $N(u)$, there is no $K_3$ formed by any three odd vertices in $N(u)$.
Let $x$, $y$, and $z$ be any three odd vertices in $N(u)$.
Assume without loss of generality that $x=P_x^i\overline{u}_i (u_{i-1}) S_x^i$, $y=P_y^j\overline{u}_j (u_{j-1}) S_y^j$,
and $z=P_z^k\overline{u}_k (u_{k-1}) S_z^k$ with $i>j>k$, where $P_x^i=P_u^i, P_y^j=P_u^j, P_z^k=P_u^k$, $S_x^i\sim S_u^i,
S_y^j\sim S_u^j$, and $S_z^k\sim S_u^k$. Note that the bits in the parentheses will appear only when its corresponding index, i.e., $i, j$, or $k$, is odd.

We will show that it is impossible that there is a $P_4$ from $x$ to $y$ and another $P_4$ from $x$ to $z$.
We only consider the case where $u_1u_0=01$. The other case, i.e., $u_1u_0=11$, can be handled similarly.

If $u_1u_0=01$, then $x_1x_0=y_1y_0=z_1z_0=11$. Suppose to the contrary that there exit paths $P_4$ in $\text{CQ}_n$ from $x$ to $y$ and from $x$ to $z$.
Let $x,v,w,y$ be the $P_4$ from $x$ to $y$. Note that the rightmost two bits of $v$ and $w$ must be in the set $\{01,10\}$.
By Lemmas~\ref{lm:lemma10} and \ref{lm:lemma11}, we have that $v_1v_0=w_1w_0=10$.
This implies that $P_x^0=P_v^0$ and $P_y^0=P_w^0$. Furthermore, vertex $v$ is adjacent to $w$ along dimension $i$.
By using a similar argument, if there is a $P_4$ from $x$ to $z$, then it would be $x,v,s,z$ with $v_1v_0=s_1s_0=10$ and $x_i=\overline{u}_i=\overline{z}_i$.
Moreover, $v$ is adjacent to $s$ also along dimension $i$, a contradiction.
Thus, in the $P_4$-graph of $N(u)$, there is no $K_3$ formed by any three odd vertices in $N(u)$ and the lemma follows.
\qed

\begin{lemma}\label{lm:lemma13}
If $\phi$ is an automorphism of $\text{CQ}_n$ for $n\geqslant 5$,
then $\phi$ maps odd vertices to odd vertices and even vertices to even.
\end{lemma}
\pf By Lemmas~\ref{lm:lemma9} and \ref{lm:lemma12}, the $P_4$-graphs of even and odd vertices
are not isomorphism. Thus it is impossible to map an even vertex to an odd vertex by $\phi$
and the lemma follows.
\qed

\begin{lemma}\label{lm:lemma14}
Let $\phi$ be an automorphism of $\text{CQ}_n$ for $n\geqslant 5$. If $\phi(u) = v$ for $u,v\in V(\text{CQ}_n)$,
then $\phi$ maps the $k$th neighbor of $u$ to the $k$th neighbor of $v$ for $k\in \{0,1\}$.
\end{lemma}
\pf Let $u=xu_1u_0$ and $v=yv_1v_0$, where $x=u_{n-1}\ldots u_2$ and $y=v_{n-1}\ldots v_2$.
First, we consider the case where $u_1u_0=00$. That is, vertex $u$ is an even vertex.
By Lemma~\ref{lm:lemma13}, vertex $v$ is also an even vertex, namely $v_0=0$.
If $v_1=1$, then we can apply $f_1(x)$ for all $x\in V(\text{CQ}_n)$ after $\phi$ is applied.
By Lemma~\ref{lm:lemma1}, the new automorphism preserves the $k$th neighbor for $k=1$ if and only if $\phi$ does.
So, we can assume $v_1v_0=00$. Accordingly, we have $\phi(u)=\phi(x00)=y00=v$.
To prove that $\phi$ preserves the $k$th neighbor for $k\in\{0,1\}$,
it suffices to show that $\phi(x01)=y01$, $\phi(x10)=y10$, and $\phi(x11)=y11$.

Now we show that $\phi(x01)=y01$. Since $x01$ is an odd vertex, by Lemma~\ref{lm:lemma13}, vertex $\phi(x01)$
is an odd vertex, namely, its $0$th bit is $1$. Since there is an edge between $x00$ and $x01$,
there is also an edge between $\phi(x00)$ and $\phi(x01)$. Recall that $\phi(x00)=y00$.
Thus $\phi(x01)$ is adjacent to $y00$. This results in $\phi(x01)=y01$.

Next we prove that $\phi(x11)=y11$ and $\phi(x10)=y10$. By Lemma~\ref{lm:lemma13}, vertex $\phi(x11)$ is an odd vertex, i.e., its $0$th bit
is $1$. Furthermore, vertex $\phi(x11)$ is a neighbor of $\phi(x01)=y01$ since  there is an edge between $x11$ and $x01$.
Thus $\phi(x11)=z11$, where $z$ is a binary string of length $n-2$. Since $\phi(x11)=z11$ is adjacent to $\phi(x10)=z10$
which is an even vertex and a neighbor of $y00$, it is impossible that $z\neq y$. For otherwise, $z10$ is not adjacent to $y00$, a contradiction.
This yields $\phi(x11)=y11$ and $\phi(x10)=y10$.
This establishes the proof of the lemma.
\qed

\begin{lemma}\label{lm:lemma15}
For $n\geqslant 5$, if $\phi$ is an automorphism of $\text{CQ}_{n+2}$, then $\hat{\phi}$ is an automorphism of $\text{CQ}_n$, where
$\hat{\phi}(v)=\lfloor\frac{\phi(4v)}{4}\rfloor$ for $v\in V(\text{CQ}_n)$.
\end{lemma}
\pf By Lemma~\ref{lm:lemma14}, it is straightforward to check that $\hat{\phi}$ is a bijection.
Moreover, if vertices $x$ and $y$ are the $k$th neighbors to each other in $\text{CQ}_n$,
then $x00$ and $y00$ are the $(k+2)$th neighbors in $\text{CQ}_{n+2}$. By Lemma~\ref{lm:lemma14} again,
vertices $\phi(x00)$ and $\phi(y00)$ are higher $(\geqslant 2)$ neighbors to each other in $\text{CQ}_{n+2}$.
Consequently, vertices $\frac{\phi(x00)}{4}$ and $\frac{\phi(y00)}{4}$ are neighbors to each other in
$\text{CQ}_n$. Therefore, function $\hat{\phi}$ preserves adjacency.
\qed

\begin{theorem}\label{thm:theorem2}
For $n\geqslant 3$, $\text{Orb}(\text{CQ}_n)=2^{\lceil\frac{n}{2}\rceil-2}$.
\end{theorem}
\pf By Lemma~\ref{lm:lemma8}, it follows that $2^{\lceil\frac{n}{2}\rceil-2}$
is an upper bound of $\text{Orb}(\text{CQ}_n)$ when $n\geqslant 3$. It remains to show that
$2^{\lceil\frac{n}{2}\rceil-2}$
is also a lower bound of $\text{Orb}(\text{CQ}_n)$ when $n\geqslant 3$.

Let $n=2k+3$ (respectively, $n=2k+4$) for $k\geqslant 1$ when $n$ is odd (respectively, even).
We claim that any automorphism $\phi$ preserves even bits $2i$ for all $0\leqslant i<k$.
By Lemma~\ref{lm:lemma13}, the claim holds when $i=0$. By applying Lemma~\ref{lm:lemma15} $i$ times for $0\leqslant i<k$
and then by Lemma~\ref{lm:lemma13}, we can find that bit $2i$ is preserved under automorphism $\phi$.
So vertices with different bits in any one of $2i$ for $0\leqslant i<k$ are in different orbits.
This further implies that $Orb(\text{CQ}_n)\geqslant 2^{\lceil\frac{n}{2}\rceil-2}$.
By Lemma~\ref{lm:lemma8}, this yields $Orb(\text{CQ}_n)=2^{\lceil\frac{n}{2}\rceil-2}$ for $n\geqslant 3$ and the theorem follows.
\qed

\section{\hspace{-0.5cm}. Concluding remarks}
\label{conclusion}

In this paper, we derive the orbit number of crossed cubes. There are a lot of variants of hypercubes, e.g.,
folded cubes \cite{Eiam91}, twisted cubes \cite{Esfa91}, m\"{o}bius cubes \cite{Cull95}, etc.
It is interesting to investigate the orbit number of those hypercube-like interconnection networks.

\end{document}